\documentclass[aps,prl,preprint,groupedaddress]{revtex4}

\usepackage{amssymb,amsmath,amsbsy}
\usepackage{mathrsfs}
\usepackage[dvips]{graphics}
\usepackage{epsfig}

\def\be{\begin{eqnarray}}
\def\ee{\end{eqnarray}}
\def\bea{\begin{eqnarray*}}
\def\eea{\end{eqnarray*}}

\def\tilde{\widetilde}

\def\centeron#1#2{{\setbox0=\hbox{#1}\setbox1=\hbox{#2}\ifdim
\wd1>\wd0\kern.5\wd1\kern-.5\wd0\fi
\copy0\kern-.5\wd0\kern-.5\wd1\copy1\ifdim\wd0>\wd1
\kern.5\wd0\kern-.5\wd1\fi}}
\def\ltap{\;\centeron{\raise.35ex\hbox{$<$}}{\lower.65ex\hbox{$\sim$}}\;}
\def\gtap{\;\centeron{\raise.35ex\hbox{$>$}}{\lower.65ex\hbox{$\sim$}}\;}

\newcommand{\newc}{\newcommand}
\newc{\qbar}{{\overline q}}
\newc{\Kahler}{K\"ahler }
\newc{\deltaGS}{\delta_{\rm GS}}
\newcommand{\SmMET}{E_T\hspace{-0.230in}\not\hspace{0.18in}}

\begin{document}
\preprint{
\vbox{\vspace*{2cm}
      \hbox{December, 2016}
}}
\vspace*{3cm}

\title{Antisplit Supersymmetry}
\author{Linda M. Carpenter}

\affiliation{Department of Physics   \\
   and \\
   Center for Cosmology and AstroParticle Physics, CCAPP \\
   Ohio State University, Columbus, OH U.S.A. \\
   lmc@physics.osu.edu \\
\vspace{1cm}}

\begin{abstract}
I explore the phenomenology of Supersymmetry models in which the gauginos are much heavier then that scalar particles of the MSSM.  In these models, the gauginos are inaccessible to colliders while the scalar spectrum is compressed.  I give several examples of models which exhibit this phenomenology built in the class of General Gauge Mediated Models.  I explore possible LSP and NLSP candidates in these scenarios including Higgsino, stau, and sneutrino candidates.   Collider signatures for these models include  multi-particle decay chains, many taus in the final state, and possible displaced vertices with semi-long lived sparticles.  I enumerate the most likely collider smoking gun collider signatures for each general LSP/NLSP scenario.

\end{abstract}


\maketitle

\section{Introduction}

Supersymmetry is among the most compelling possibilities for Beyond the Standard Model physics.  It offers a natural solution to the hierarchy problem, the possibility of the unification of forces, Dark Matter candidates, and a panoply of new particles which may be discovered at the scale of current colliders.

The low energy spectrum of the MSSM is, of course, completely dependent upon the mechanisms of the communication of SUSY breaking.  Simple and elegant SUSY breaking schemes exist which are predictive, containing only a small number of new parameters, as well as being flavor blind.  Among these, for example are Anomaly Mediation or Minimal Gauge Mediation\cite{Dine:1993yw}\cite{Dine:1994vc}\cite{Dine:1995ag}.  Many minimal mass giving mechanisms  produce spectra with a the uniform feature that scalars and gauginos have roughly equivalent masses.  In addition particles charged under QCD are generally heavy, and other particles light.  A typical ratio between the squarks and the light neutralinos, for example is 8 to 1.  Experimental signatures of such models, while having some special identifying features, show a large degree of uniformity.  Among the most prominent signals of such models for example is jets plus missing energy.  The expectation is that the gluinos will be among the first SUSY particles to be observed, with a large production cross section, and that they will decay through on or offshell quarks down to the LSP which is most canonically the lightest neutralino.

It is possible to generate MSSM spectra which are extremely divergent from minimal predictions.  Slightly more general models allow for a significant  departures from standard collider signals.  For example, a generalized definition of Gauge Mediation  is simply that the MSSM particle masses go to zero if the standard model gauge couplings are turned off \cite{Meade:2008wd}.  There are a large variety of models that fit this definition.  In the original incarnation of General Gauge Mediation, the MSSM spectrum depended on six mass parameters and satisfied certain sum rules.  There are three separate mass parameters for gauginos and three more for scalars, one each for the SU(3), SU(2) and U(1) components of the fields.  In principle, the masses of the gauginos and the scalars are completely independent in these models, and needn't follow any standard pattern.  This incarnation of General Gauge mediation may be even further generalized, allowing different operators which allow for modified sum rules and possibly adding even more parameters in the spectra, for example A terms \cite{Intriligator:2010be}.

A variety of SUSY models exist which produce highly non-standard spectra, and many of the interesting features of these models come from introducing a hierarchy between the scalar and gaugino masses.  In Split Supersymmetry , the gauginos are accessible to colliders, while the scalars are extremely heavy - producing signatures such as extremely long lived supersymmetric particles \cite{ArkaniHamed:2004fb} .  Supersoft Supersymmetry is a mechanism with extra chiral adjoint fields in which the gauginos are around a TeV in mass, and the scalars are a square root of a loop factor below the gaugino mass \cite{Fox:2002bu}.  There is also the Higgsino World and its variants, in which all superpartners are multi TeV and only the Higgsino-like gauginos and Higgs sector scalar  fields are light and accessible to colliders \cite{Baer:2011ec} .

I propose to analyze here the phenomenology of a general class of models I refer to, with deference to the source material,  as Antisplit supersymmetric models.  These are models in which the gaugino masses are much heavier than the scalar sparticle masses.  Exemplary models in this class include Gaugino Mediation\cite{Chacko:1999mi} \cite{Kaplan:1999ac} and R-symmetric models in which gauginos are Dirac instead of Majorana particles \cite{Hall:1990hq}.  This spectrum, however can also be found in General Gauge Mediated models and models with hybrid gravity-gauge mediation as I will discuss later in this article.  I will introduce models which maximize the gaugino-scalar mass splitting.  In particular, I will here study models where gauginos are of the scale a few to 10 TeV and thus inaccessible to colliders, while the scalar superpartners remain weak or TeV scale with a compressed mass spectrum.   The phenomenology of these models is quite different from that of standard MSSM spectra.  One can expect several interesting features.  First, gaugino mass parameters are quite large, of order 10 TeV, but the mu term remains weakscale, thus the lightest neutralino and charginos are Higgsino-like. As has been pointed out in ref \cite{Kribs:2012gx} regarding Supersoft models, since gauginos are very heavy, the production cross section of pairs of MSSM sfermions-for example squarks- is much lower than in Standard SUSY scenarios.  This means that mass limits on squarks and sleptons from standard searches like jets plus missing energy, are quite a bit lower than other SUSY scenarios, even assuming quite simple decay scenarios for scalars.   Further, the scalar decay chains may become quite non-standard,  involving decays through on or offshell Higgsinos, often leading to heavy flavor fermions as decay products. The phenomenology of such spectra has been explored only in a piecemeal fashion such as \cite{DeSimone:2008gm}.  If the SUSY breaking scheme is Gauge Mediation, we can expect that NLSP to decay, perhaps inside the detector and possibly with a displaced vertex.  The LSP candidates I will consider here are  stau, Higgsino, or sneutrino. Decays of NLSPs can put interesting particles in the final state such as taus, Higgses or Z bosons.  Standard Gauge Mediation with bino-like NLSP produces photons in the final state, but Higgsino-like gauginos favor other final states.   I present below an illustration of a typical Antisplit mass spectrum

\begin{figure}[h]
\centerline{\includegraphics[width=13 cm]{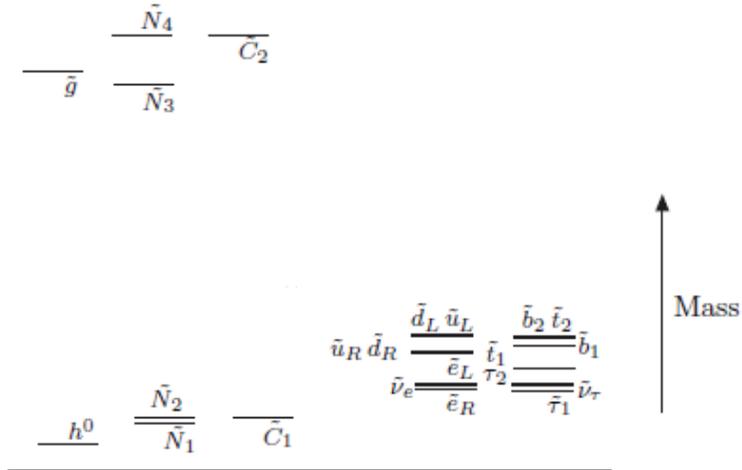}}
\caption{{Illustration of  a typical Antisplit SUSY spectrum with extremely heavy gauginos and  light scalars.}}
\label{fig:2pgm}
\end{figure}

Though it may seem that getting a spectrum in which gauginos are extremely heavy is theoretically challenging, it follows quite naturally by modifying certain existing models of SUSY breaking communication.  I will give several examples of implementations  SUSY breaking which achieve an Antisplit spectrum.  The achieve this spectrum I will take advantage of a special version of Gauge Mediation which allows negative R-symmetric contributions to scalar mass squareds which do not contribute to gaugino masses.

In order to demonstrate  that this general spectrum can be achieved with many forms of SUSY breaking communication, I will discuss two models of a generalized General Gauge Mediation.   One model is simply one in which gauginos have normal gauge mediated masses set to the multi TeV scale.  The scalar mass contributions are then canceled by gauge mediated masses which follow from additional messengers with negative supertrace.  The gaugino masses may either be Majorana or Dirac, however the choice of Dirac masses for gauginos means tuning in the model is substantially less.  I will outline both typed of models and calculate example a mass spectra based on the Dirac-gaugino model. In this set of models, the gauginos acquire Dirac masses by marrying fields which are adjoint under the Standard Model gauge groups.  The scalars then get a finite mass contribution due to loops of the scalar adjoint which is the square root of a loop factor below the gaugino masses.  This is the building block of a scheme where the gauginos are naturally larger than the scalars.  By adding to this scheme additional messengers with negative supertraces  the result is a modified Supersoft model in which I can make the gauginos arbitrarily large compared to the scalars.  I will then explore the phenomenology of models with these spectra. I will produce example spectra for the cases of stau, Higgsino, and sneutrino NLSPs.  I will then enumerate the most probably decay chains for these particles with these mass spectra, and discuss the final state topologies most likely to be discovery scenario for this type of model.

This paper proceeds as follows, section 2 reviews general gauge mediation and presents several model-types capable of producing an Antisplit supersymmetric spectrum, section 3 presents the phenomenology of a stau NLSP in this scenario, section 4 presents Higgsino lsp/NLSP phenomenology while section 5 present sneutrino LSP/NLSP phenomenology.  Section 6 concludes.

\section{Models}

In General Gauge Mediation the minimal mass-relations between gauginos and scalars do not hold.  Gaugino masses are independent of each other and scalars may get mass independent of the gauginos.  This may be achieved by including multiple sets of messengers, including multiple SUSY breaking spurions, and/or having several distinct SUSY breaking sectors \cite{Buican:2008ws}\cite{Carpenter:2008wi}.    I will present below two simple models which are generalized Models of Gauge Mediation capable of producing spectra in which the gauginos are much heavier than the scalar particles of MSSM.
\subsection{A Model with Dirac Gaugino}
 Supersoft Mediation is a very simple and interesting Mediation Mechanism which naturally has light scalars compared to gauginos.  In this scenario, the gauginos get large Dirac masses arising from the coupling of the gauginos to new chiral adjoints and a hidden sector $U(1)^{'}$ gauge field which gets a D term.  In the low energy, one may write a superpotential term
 \be
 W = c_i\frac{W^{'}W_i A^i}{\Lambda}
 \ee
 where $W^{'}$ is the hidden sector U(1) gauge field and  A is the adjoint.  The index i runs over the three SM gauge groups.  Here the gauge indices are contracted between $W$ and the adjoint while Lorentz indices are contracted between the W's. Once the $U(1)^{'}$ D term is inserted, the operator becomes a Dirac mass for the gauginos

 \be
c_i \frac{D}{\Lambda}\lambda_i \psi_{Ai}
 \ee
which has the value $c_i{D}/{\Lambda}$.  All of the gauginos may have independent mass parameters as the coupling $c_i$ may vary for each gaugino.  One can imagine generating the SUSY breaking D term by employing any number of dynamical models \cite{Csaki:2013fla}, or simpler O'Raigherty-like models in which U(1) is the only gauge group in the DSB sector \cite{Dine:2006xt}  \cite{Carpenter:2010as}.

It is simple enough to ensure that the gaugino masses arise from a General Gauge Mediated mechanism.  One can generate the above coupling of  adjoint fields the MSSM gauginos by employing  a set of Messenger fields, which are charged both under the standard model gauge groups and the new U(1) gauge symmetry.  One may write a superpotential for messenger fields, here called T.

\be
W_T= m_T T\overline{T}+ y_i\overline{T}AT
\ee
where the messengers have a supersymmetric mass term. A one loop mass diagram for gauginos is generated resulting gaugino a gaugino mass term

\be
m_{\lambda_i} = \frac{g_i}{16\pi^2}\frac{y_i D}{m_T}
\ee

Note that this mass term does indeed comply with the definition of a General Gauge Mediated model, that is, the gaugino mass goes to zero if the standard model gauge coupling is turned off.

Only after the gaugino has gotten mass are the scalar mass-squared's generated.  These masses are finite and one loop level lower than the gaugino masses, involving the gauginos or real scalar adjoint in the loop.  The scalar masses from the Supersoft process are
\be
m_s^2= \frac{C_i \alpha_i {m_{\lambda i}}^2}{\pi} \log (\frac{\delta_i }{ m_{\lambda i}})^2
\ee
where $m_{\lambda i}$ are the gaugino masses , $\delta_i$ is the mass squared of the real part of the adjoint field and $C_i$ are the Casimirs of the fields.  One can see that in this model the scalars are naturally much lighter than the gauginos,  the ratio between MSSM scalar and gaugino masses is
 \be
 \frac{m_s}{m_\lambda}= \sqrt{\frac{2 C_i\alpha_i }{\pi}\log (\frac{\delta_i }{ m_{\lambda i}})}
 \ee
 Under normal circumstances, the value of the Log in the above equation is 4, as the real part of the scalar adjoint is twice the gaugino mass. We then we expect 10 TeV gauginos to yield roughly 1 TeV scalar particles.  However, if $\delta_i$ equals $m_{\lambda i}$, the gaugino masses make no contribution Supersoft scalar masses at all.  As one changes the value of this log, the gauginos may be made arbitrarily heavier than the scalars.  Indeed both the scalar and pseudo-scalar adjoint fields receive corrections to their SUSY breaking masses through loop effects.  The value of these corrections depends on the details of the model.  For example, in gauge mediated implementations, the value of these operators will be determined by the specific number and masses of the messenger fields T.  Discussions of these operators may be found, for example, in     \cite{Csaki:2013fla}   \cite{Carpenter:2010as} \cite{Carpenter:2015mna} and \cite{Nelson:2015cea}.

One very simple way to achieve negative scalar mass contributions is though an R symmetric process.  One adds to the model a new set gauge messengers with non-holomorphic masses.  These then create two loop log divergent scalar masses which are  proportional to the messenger supertrace.  These mass contributions were first calculated by calculated in \cite{Randall:1996zi}\cite{Poppitz:1996xw} and are given by

\be
m^2_i = - \sum_a\frac{g_a^4}{128 \pi^4} S_Q C_{ai} {\rm Str} M_{mess}^2 \log(\frac{M^2}{\Lambda^2})
\ee
where S is the Dynkin index of the messengers, $C_{ai}$ is the Casimir for the scalars, and M is the supersymmetric messenger mass.  These masses have several interesting features, first note they are slightly UV sensitive as they are log divergent.  Second, the physics involved in generating these masses is R symmetric, it contributes to the scalar mass squareds without changing the gaugino masses.  Finally,  the scalar mass squareds have signs which are the opposite of the
sign of the messenger supertrace; therefore messengers with positive supertrace will yield the negative mass contributions desired for adjoints without changing the gaugino masses.  The messengers may participate in a two tiered system where themselves receive non-zero supertraces from gauge mediation as in  \cite{Randall:1996zi}, but they may perhaps get supertraces in  some form of direct mediation, where the messengers themselves participate in SUSY breaking \cite{Poppitz:1996fw} .

The spectra generated for such a model are quite interesting.  First there are three independent gaugino masses given by eqn. 4.  The scalar masses are given by two contributions,

\be
m_s^2={m_{Supersoft}}^2+{m_{SuperTrace}}^2
\ee
the supertrace mass term is given by eqn. 7, it is a contribution felt by all scalars, including scalar adjoint fields, as it arises due to the presence of a new set of gauge messengers charged under the standard model gauge groups.  Here the Supersoft mass contributions to the MSSM scalars(excluding the adjoints) is given by
\be
{m_{Supersoft}}^2= \frac{C_i \alpha_i {m_{\lambda i}}^2}{\pi} \log (\frac{\delta_i }{ m_{\lambda i}})^2
\ee

where

\be
\delta_i^2= (2\lambda_i)^2- \sum_a\frac{g_a^4}{128 \pi^4} S_Q C_{ai} {\rm Str} M_{mess}^2 \log(\frac{M^2}{\Lambda^2})
\ee

is the mass squared of the real scalar adjoint fields, which itself gets mass contribution from additional supertrace-ful messengers.  In the above equations additional contributions to the scalar adjoint mass are neglected. They themselves will not effect the spectrum too much as they are expected, at most, to be of the size $m_{\lambda}^2$, if the messenger sector has been so arranged to ensure their positivity.   We can see that the addition of new messengers, themselves with positive supertrace, may subtract substantially from MSSM scalar masses and drastically alter the scalar/gaugino mass ratio in eqn 6.    In order to make a substantial mass correction, the Supersoft and supertrace terms in eqn. 8 must cancel at order 1.  In order to make a mass contribution of order weak scale, we should demand that the messenger supertrace should be in the region of 10 to 100 TeV.

\subsection{Models with Majorana Gauginos}

A second method for generating an anti-split spectrum follows as a simple modification to normal gauge mediated models with Majorana gaugino masses.  I present here a model which is quite simple, but very tuned.

The simplest incarnations of gauge mediation require only the presence of messengers charged under the standard model gauge groups, which also couple to a SUSY breaking spurion.    The spurion has both a vev and an F term, generating a supersymmetric and non-supersymmetric mass for the messengers, one can write the messenger superpotential
\be
W = X M \overline{M}\rightarrow v M\overline{M} + \theta^2 F_X M\overline{M}
\ee
where X is the SUSY breaking spurion and M are the messengers,in a fundamental and anti-fundamental of the SM gauge group SU(5).  Majorana gaugino masses are generated though a one loop process involving the messengers.  The resulting gaugino mass is
  \bea
  m\lambda_i = \frac{\alpha_i}{4 \pi}F/v.
   \eea

Note that in this case all gaugino masses are proportional to a single mass parameter.  If one employs multiple SUSY breaking spurions and several sets of messengers whose SU(2) and SU(3) components couple differently to the spurion, similar one loop messenger diagrams can generate three distinct mass parameters for gauginos as in \cite{Carpenter:2008wi}.

Generating multi-TeV gaugino masses in this scenario simple requires the correct choice of the parameters $F/v$. The problem with the spectrum is that scalar masses are generated at two loop order and are roughly of equivalent size the gaugino masses,

\be
m_{s}^2 \sim \frac{\alpha_i}{4\pi}^2 F^2/v^2
\ee

However one can see from the previous subsection, that it is possible to generate R symmetric negative contributions to scalar mass squareds by adding messengers with a positive supertrace.  Thus, to get an Antisplit mass spectrum  easily and without the addition of extra adjoint fields, one may simply cancel the normal gauge mediated scalar masses against the R symmetric gauge mediated scalar masses.
\be
m\lambda_i = \frac{\alpha_i}{4 \pi}F/v
\ee

\be
m_{s}^2 \sim \frac{\alpha_i}{4\pi}^2 f^2/v^2 - \sum_a\frac{\alpha_i}{4\pi^2}^2 S_Q C_{ai} {\rm Str} M_{mess}^2 \log(\frac{M^2}{\Lambda^2})
\ee

In this case, if one wished to generate 10 TeV gauginos, then two loop gauge mediated diagrams will also generate 10 TeV scalars.  Canceling the scalar masses down to the weak scale involves tuning the two mass squared parameters to cancel to one part in ten thousand, a simple but uncomfortable prospect.  This would also require a fairly large messenger supertrace and would favor a higher energy SUSY breaking scale than the previous example.

\subsection{Deflected Gravity Mediation Models}
Another quite simple way to get a spectrum in which the gauginos are heavy compared to the scalars is to use the gravity mediated formalism. This, most simply involves MSUGRA.   Here, one usually gives the scalars and the gauginos universal masses at a very high scale. The masses are then run down to a the collider scale and the mass differences observed in the spectrum occur only due to running.  The $\mu$ term is a parameter picked by hand.  This can always be adjusted to ensure that the Higgsino is much lighter than the gluino/weakinos. Some of the work achieving an Antisplit SUSY spectrum can be done by adjusting the ratio of the  scalar mass $m_0$ to the gaugino mass parameter $m_{1/2}$.  One may also effect the phenomenology by choosing the value of the high energy scale, which controls how long the mass parameters are run.  This alone, however is not enough.  To achieve true flexibility of the spectrum one can add a bit of general gauge mediation, namely, the same negative mass contribution as the models above, gotten adding additional messenger fields with positive supertrace.   In principle these type of messengers can also be added to  Deflected Mirage Mediation \cite{Everett:2008qy}.  In which one combined the mediation mechanisms of Gravity Mediation, Anomaly Mediation and normal Gauge Mediation.  In standard Deflected Mirage Mediation one picks universal scale $m_0^2$ from which sparticle masses evolve from the combination of gravity and anomaly mediation. In addition one adds sets of gauge messengers with an F term mass and supersymmetric mass v, from the vev of a dynamical field.  One sets the scale of F/v just like in normal gauge mediation and this deflects the SUSY mass spectrum.   In principle, one could instead add R-symmetric gauge mediation discussed above where messengers only get a non-holomorphic mass from SUSY breaking.  One adds sets of messengers with supertraces $STr M_i^2$   the index i indicating SU(3), SU(2) or U(1) messengers at some arbitrary SUSY breaking scale.  As the theory runs this will generate negative mass squared contributions to the MSSM scalar particles.  This will subtract arbitrary mass contributions to the MSSM scalars charged under the three gauge groups and many variant low energy spectra may be achieved.

The decay chains in the case of an deflected MSUGRA scenario are quite different than in the general gauge mediated scenarios as in MSUGRA, the gravitino is not the LSP.  In gauge mediation, the NLSP may decay to its superpartner and a gravitino inside the detector, which adds states to the decay chain.   In the deflected MSUGRA scenario however, the LSP may not be a stau but must be a neutral particle.

In the phenomenology section below, I am mostly concerned with gauge mediated signatures of Antisplit SUSY and most of the discussion will involve the decay of the NLSP inside the detector.

\subsection{Phenomenology}
Below I will describe the phenomenology of several benchmark spectra of Antisplit models, classifying the models by LSP or NLSP as the case may be.  General Gauge Mediated scenarios allow for a wide variety of NLSP candidates \cite{Carpenter:2008he}\cite{Grajek:2013ola}.   In Gauge Mediated models the gravitino acquires a mass that depends on the SUSY breaking F term $m_{\tilde{G}} = F/\sqrt{3}M_{P}$. There is a universal coupling of sparticle to their particle partners and to the gravitino.  This decay width is suppressed by powers of the SUSY breaking scale F.  For example the decay width  of a sfermion into fermion and gravitino is
\be
\Gamma_{\tilde{f}}=\frac{1}{16 \pi }\frac{m_f^5}{F^2}(1-\frac{m_G^2}{m_f^2})^4
\ee
In the case of low energy SUSY breaking, the gravitino is quite light and the NLSP will decay to the gravitino in short order. The limit for decay length $c\tau$ inside the detector places an upper bound on the F term of roughly $10^6$ GeV.   We may demand that the sparticle decay take place inside the detector, for Gauge Mediated scenarios and demand that for Antisplit models the gauginos lie in the multi to 10 TeV range. This then limits the messenger mass M are at quite a low scale, not above $10^7$ GeV with the F term not much below.  For scenarios where the NLSP does not decay inside the detector, we may probe higher SUSY breaking and messenger scales.    Prospects for SUSY breaking in this scenario are quite interesting.  First, for GGM models with prompt NLSP decay, we are impinging on the perturbative requirement that $\sqrt{F}<v$.  This brings us into the realm of direct mediation, where the messengers themselves participate in SUSY breaking.  Models with Dirac gauginos often require that SUSY breaking comes from a D term.  However this is a field dependent D term, which comes from F-term SUSY breaking, often at a higher scale. It is, however possible to choose a SUSY breaking sector such that the D-term and the F-terms are roughly of the same order.  This occurs, for example in the 4-1 model.  This is, however, a further indication that the SUSY breaking sectors required for low scale F terms are complex.  A second note is here in order, these models naturally demand an F term which is just on the edge of NSLP decay inside the detector. This means that the models discussed naturally lend themselves to the NLSP decaying with $c\tau$ in the mm to meter range, leaving a striking signal of displaced vertices and long lived particles.

In the case that the sparticle cannot decay to a final state gravitino inside the detector or otherwise, the set of LSPs is constrained to be neutral.  Stable charges particles are would cause a cosmological catastrophe, in addition collider bounds on neutral particles are quite weak. Therefore I consider in this work relevant stable LSP/NLSPs to be neutral, either the lightest neutralino or the sneutrino.  In the case that the lightest sparticle is unstable, and gauge mediation obtains, the NLSP may be charged, as it decays to a final state charged standard model particle and gravitino.  I will consider in this case the set of NLSPs to be the $\chi_0$, $\tilde{\tau} $, and $\tilde{\nu}$.  In creating supersymmetric spectra for these cases, I have pulled benchmark points from the modified Supersoft model in the second section.  I chose this model as it was the least tuned and most phenomenologically viable.  An interesting point about these models is that they may be completed with a number of new Higgs sector operators which aid in electroweak symmetry breaking, provide mu terms and substantially raise the Higgs mass even for lower values of $tan\beta$ and lighter stop masses \cite{Nelson:2002ca}\cite{Benakli:2012cy}.

\section{Stau NLSP}

I consider here the phenomenology of a gauge mediated model with a stau NLSP. In such a model the stau is the final sparticle in the MSSM decay chains, which then decays promptly to a gravitino and a tau.  I have created an example MSSM mass spectrum using the modified Supersoft model.  Here I have chosen gaugino mass parameters $M_1$=5TeV $M_2$=$M_3$=6TeV.  The Supertrace of the additional messengers was chosen in the 10 TeV range.   The spectrum appears below in table 1.

\begin{table}[h]
\begin{center}
\begin{tabular}{|c c|c c|c c|c c|}
\hline
neutral gaugino &  mass & chargino & mass & squark & mass & slepton & mass  \\
\hline
 $m_g$  & $6000$ &   &  & $m_{u_l}$ & 903 &  $m_{e_l}$ & 463  \\
 $m_{\chi_4^{0}}$ & $6001$ & $m_{\chi_2^{+}}$ & $6001$  & $m_{d_l}$ & 903 & $m_{e_r}$ & 203 \\
 $m_{\chi_3^{0}}$ & $5000$  &   &  &  &   &  &  \\
 $m_{\chi_2^{0}}$ & $500.4$  & $m_{\chi_1^{+}}$ & $499.5$ &   &  &  & \\
 $m_{\chi_1^{0}}$ & $498.9$  &   &  & $m_{u_r}$ & $782$ & $m_{\tau_1}$ & 199  \\
 & & & &  $m_{d_r}$ & $784$  &   &  \\

\hline
\end{tabular}
\label{fields}
\end{center}
\end{table}

Here the $\mu$ term has been chosen to be 500GeV with a modest $tan\beta$ of 3.

Notice that  I have chosen a point where the right handed sleptons are lighter than the left handed sleptons,  hence the right handed sfermions are lighter than the left handed.   This however is an arbitrary choice, and once could easily have chosen the left handed states to be lighter. The right handed sleptons are the lightest fields and due to running, the NLSP is the right handed stau.
The spectrum has several features which are common to Antisplit models.  First of all, the gluino and heavier chargino and neutralinos are too heavy to be observable  by LHC.   The lighter neutralinos and chargino are Higgsino like which  means that their masses are quite degenerate.  In addition the scalar mass spectrum is compressed.  The squarks are quite light, however given their departure from standard decay chains this compressed non-standard SUSY scenario is not yet harshly constrained by LHC.  In fact, according to initial Supersoft studies, pair production of squarks with 800 GeV mass is only 10fb in the 8 GeV run of LHC, slightly scaling up results in figure 3 of reference  \cite{Kribs:2012gx}.

The branching fractions for this spectrum were calculated using the SUSYHIT\cite{Djouadi:2006bz} and the BRI routine of the BRIDGE program \cite{Meade:2007js}, I have listed the most relevant decays of the above spectrum in table 2.

\begin{table}[h]
\begin{center}
\begin{tabular}{|c c c c| c c c|}
\hline
particle &  decay & b.f.  &  & particle & decay & b.f. \\
\hline
$\tilde{u_l}$  & $\chi^{\pm}_1 $d & .78 &  & $\tilde{t}_2$ & $\chi^{\pm}_1 $ b   & .38\\
-  & $\chi^{0}_1 $ u & .20 &   &- & $\chi^{0}_1 $ t   & .57  \\
-  &    &   &  &$\tilde{t}_1$ & $\chi^{0}_1 $ t   & .72 \\
$\tilde{u_r}$  & $\chi^{0}_1$ u & .96 &  & - &  $\chi^{\pm}_1 $ b  & .27 \\
\hline
$\tilde{d_l}$  & $\chi^{\pm}_1 $u & .49 &  & $\tilde{b}_2$ & $\chi^{\pm}_1 $ t   & .99\\
-  & $\chi^{0}_1$ d & .49 &   &$\tilde{b}_1$ & $\chi^{0}_1 $ b   & .45 \\
$\tilde{d_r}$  & $\chi^{0}_1$ d & .96   &  & - &  $\chi^{0}_2 $  & .41 \\
\hline
$\chi^{\pm}_1$ & $\nu_{\tau}$ $\tilde{\tau^{\pm}}$ & .94 & & $\tilde{\ell}_L$ & $\ell$ $\ell^{' \mp}$ $\tilde{\ell^{\pm}}_R^{'}$   & .632 \\
$\chi^{0}_1$ & $\tau^{\mp}$ $\tilde{\tau^{\pm}}$ & .79 &   & - &$\ell$ $\tau^{\mp}$ $\tilde{\tau_R^{\pm}}$   & .316 \\
- &  $\ell^{\mp} \tilde{\ell^{\pm}}$ & .17 &  & $\tilde{\ell}_R$ & $\ell$ $\tau^{\mp}$ $\tilde{\tau}_R^{\pm}$   & 1 \\
$\chi^{0}_2$ & $\tau^{\mp}$ $\tilde{\tau^{\pm}}$ & .96 &  & $\tilde{\tau_R}$ & $\tau$ $\tilde{G}$ & 1 \\
\hline

\end{tabular}
\label{fields2}
\end{center}
\end{table}

\begin{figure}[h]
\begin{center}$
\begin{array}{cc}
\includegraphics[width=3.0in]{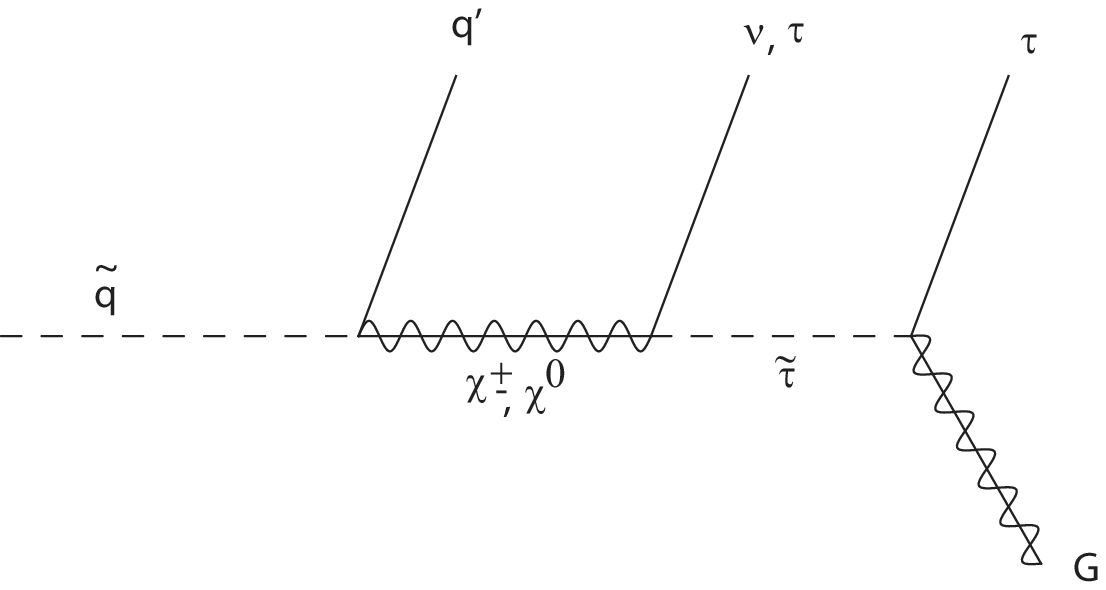} &
\includegraphics[width=3.05in]{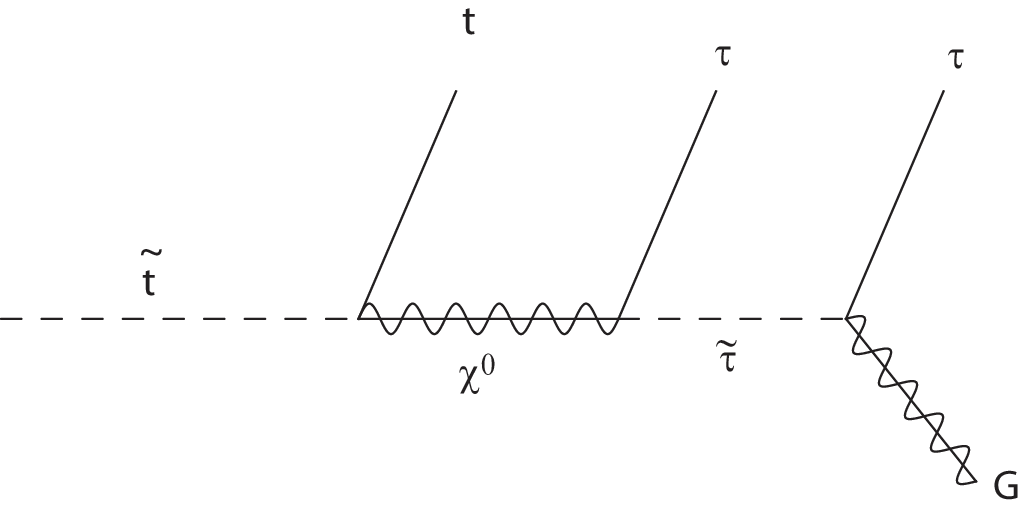}
\end{array}$
\end{center}
\caption{Typical heavy and light flavor squark decay chains in an Antisplit scenario with stau NLSP. }
\end{figure}

One sees from this table that staus appear in all decay chains.  In addition to the fact that all decay chains end in the stau NLSP decaying to a tau and gravitino,  taus are favored decay of the charginos and neutralinos which are Higgsino-like. The mixing between the Higgsinos and the gaugino-like states is quite small, of order a few percent.

In this scenario one expects that the quarks are the most abundantly produced objects.  The light $\tilde{d}$ and $\tilde{u}$ squarks decay through onshell chargino and neutralinos mostly through the chargino/neutralino mixing with the gaugino-like states due to the very small Yukawa couplings to the Higgsinos.  One notices that the right handed states prefer to decay through the neutralino rather than the chargino as they do not couple to the wino and couple very weakly to the Higgsino. The stop however have much larger coupling to the Higgsino portions of the charginos/neutralinos.  For example, right handed stops  have a larger branching fractions to the chargino than the light up squarks, as the right handed states have large coupling to the Higgsino part of the chargino admixture. I present in figure 2 typical diagrams of the light and heavy flavor squark decays.

\begin{figure}[h]
\centerline{\includegraphics[width=4 cm]{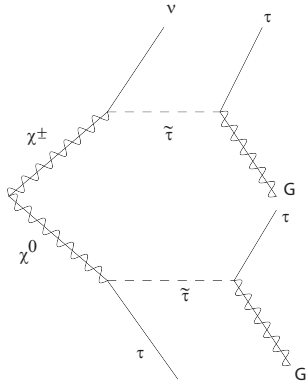}}
\caption{{Chargino Neutralino pair production with decay chains through stau NLSP.}}
\label{fig:2pgm}
\end{figure}

Notice that the decay chain of the stop is especially striking, containing multiple particles of heavy flavor.  Stop pair production would involve a final state with 2 top quarks, four taus and missing energy.  Similarly a light sbottom decay would contain bottoms, staus and missing energy.

Since the light neutralinos and charginos are Higgsino-like they overwhelmingly decay to a tau neutrino and the stau NLSP or to a tau stau pair respectively.  Collider signatures for the light neutralinos and charginos are thus quite different than the standard SUSY case where these particles are usually gaugino like and thus decay rather democratically into all flavors of lepton.  I present as an example process, the process of a charged and neutral Higgsino in figure 3.

Here the final state signature is  $3\tau+ \SmMET$.  Pair production of the charged or neutral state will result in with 2 or 4 taus plus missing energy respectively.  The multi-tau plus missing energy signature replaces the standard smoking gun tri-lepton signature for MSSM charginos-neutralino pair creation.  In addition, one expects the production cross section for Higgsino-like neutralinos to differ from that of gaugino-like fermions.

\begin{figure}[h]
\centerline{\includegraphics[width=10 cm]{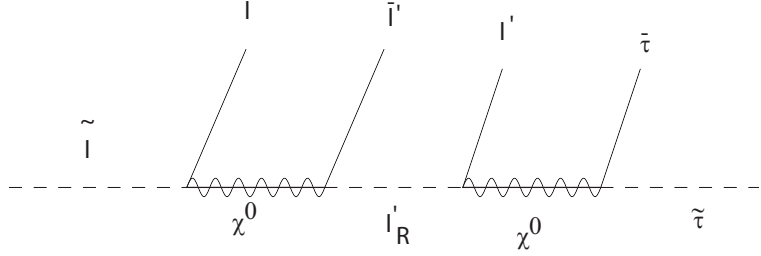}}
\caption{{Decay chain of left handed sleptons through right handed sleptons and stau NLSP.}}
\label{fig:2pgm}
\end{figure}

Finally, pairs of light flavors of sleptons may be produced from weak processes.  These decays are particularly interesting because they must also involve taus in the final state.  The decay of both the left and right handed slepton is three body decay through an offshell neutralino.  The right handed slepton undergoes three body decay to same flavor lepton, and a tau and stau pair.  The tau of the tau-stau pair may have either sign.  Its decay chain is the $\tilde{\ell} \rightarrow \ell \tau^{\mp} \tilde{\tau}^{\pm}$.  The decay of the left handed sleptons is a multi-step process, as described below.

Here we see that the left handed slepton decays first through an offshell gaugino to the right handed slepton and two hard leptons.  The initial stage of the process is $\tilde{\ell_L} \rightarrow \ell \ell^{'\mp}\tilde{ \ell^{'}}^{\pm}$  The final state slepton may be a stau or a slepton of light flavor.   In is interesting that the right handed slepton may be of any flavor.  Thus, a left handed slepton will decay with one same flavor lepton and two leptons of arbitrary flavor. The resultant right handed slepton, if not a stau,  then decays again through an off-shell gaugino as discussed above.  A final state right handed stau will always decay to a tau and a gravitino.  As an example, pair production of right handed sleptons would result in multi-lepton final states, containing 2 leptons, 4 taus and missing energy.  For extremal values of F-terms in the high energy model, the lifetime of the NLSP, the right handed stau may be appreciable and result in a displaced vertex or kink in addition to the other particles in the event.

\section{Higgsino NLSP}

It is quite easy to produce spectra in which the Higgsino is the LSP or NLSP in Antisplit scenarios. This simply involves choosing a $\mu$ term which is light compared to the other parameters.  As an example I present a benchmark point below, using the same parameters as the previous benchmark point only shifting the value of the $\mu$ term to 170 GeV.  The results of the spectrum are in table 3.  Again, a light Higgsino means that there are three degenerate states, two neutralinos and one chargino which are split by a few hundred MeV.  In some scenarios, for example in deflected gravity mediation or high scale gauge mediation, the lightest Higgsino will be absolutely stable or stable on collider timescales.  In gauge mediated scenarios, where we expect a light Higgs boson, the light neutralino decays to a Higgs or gauge boson and a gravitino.  This means that gauge mediated decay chains have the interesting  property of containing electroweak gauge bosons or Higgses in the final state. Below I give an example spectrum which had a Higgsino-like  neutralino NLSP.

\begin{table}[h]
\begin{center}
\begin{tabular}{|c c|c c|c c|c c|}
\hline
neutral gaugino &  mass & chargino & mass & squark & mass & slepton & mass  \\
\hline
 $m_g$  & $6000$ &   &  & $m_{u_l}$ & 903 &  $m_{e_l}$ & 463  \\
 $m_{\chi_4^{0}}$ & $6001$ & $m_{\chi_2^{+}}$ & $6001$  & $m_{d_l}$ & 903 & $m_{e_r}$ & 203 \\
 $m_{\chi_3^{0}}$ & $5000$  &   &  &  &   &  &  \\
 $m_{\chi_2^{0}}$ & $-170.4$  & $m_{\chi_1^{+}}$ & $169.5$ &   &  &  & \\
 $m_{\chi_1^{0}}$ & $169$  &   &  & $m_{u_r}$ & $782$ & $m_{\tau_1}$ & 199  \\
 & & & &  $m_{d_r}$ & $784$  &   &  \\

\hline
\end{tabular}
\label{fields3}
\end{center}
\end{table}

\begin{figure}[h]
\centerline{\includegraphics[width=10 cm]{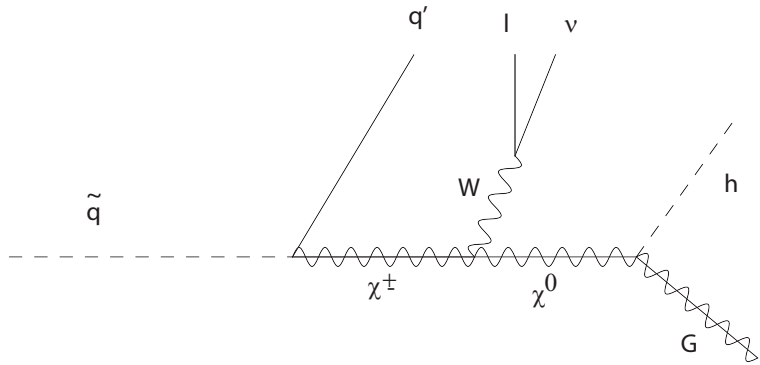}}
\caption{{Decay chain of squark through Higgsino NLSP.  Here the NLSP decays to a Higgs and a Gravitino}}
\label{fig:2pgm}
\end{figure}

The decay of the chargino and second lightest neutralino have been widely studied in cases where the light neutralinos and charginos are degenerate, for example in Anomaly Mediated scenarios where the LSP is wino like, and in the Higgsino world scenario.  Here the heavier charged and neutral sates must decay to the lightest neutralino through an off shell gauge boson.  The above point quotes a tree level result for chargino/neutralino masses. It have been pointed out that due to loop effects, the mass splitting will always be large enough to allow the chargino to decay into a neutralino and a charged pion.  This will be the most probable decay of the chargino.  However in the chargino rest frame decay products, whether a pion or charged lepton from the chargino decay will, be quite soft due to the small mass spitting.  In some cases, the chargino will be long lived, and may decay with a displaced vertex appearing as a kink or a stopped track in the tracking chamber.

In scenarios with stable neutral Higgsinos, Higgsino pair production if quite hard to observe.  The processes $pp \rightarrow \chi^{+} \chi^{-}, \chi_0 \chi_0,\chi^{\pm} \chi^{0}$ proceed through electroweak quark fusion and have weak scale production cross sections. The Higgsino decay products are quite soft and appear invisible in many searches.  One search strategy for this scenario is observing initial state radiation plus a missing energy signal in the form of a mono-jet, mono-photon or mono-Z search. However, discovery prospects for this scenario are still quite challenging with discovery potential around 200 GeV for 3 inverse attobarns of LHC data \cite{Anandakrishnan:2014exa}\cite{Baer:2014cua}.  Though the Higgsinos themselves may be hard  to observe in this scenario of stable or meta-stable Higgsino, bounds on other sparticles may be quite harsh.  For example, squarks would decay to a quark plus a chargino or Higgsino, both would likely appear as missing energy to the detector. Thus the discover signal for squarks is the simple 2 jets plus missing energy channel leading to observable squarks, though with less production cross section than in standard scenarios, as discussed above.  Slepton decays would follow a similar trajectory. In fact this scenario was studied in reference  \cite{Eckel:2014dza} and found that unless the spectrum was quite compressed, left handed slepton mass bounds might be set as high as 490 GeV.  This is because over much of parameter space charged sleptons prefer to decay to a lepton and lightest neutralino, rather than a neutrino and chargino.  Additionally, over most of parameter space sneutrinos prefer to decay to a lepton and chargino.  Constrains on searches with 2 hard leptons and missing energy thus place strong lower bounds left handed slepton masses.

In the case that the Higgsino is the  NLSP, it will eventually decay to a Higgs or gauge boson plus missing energy.  For our scenario the NLSP is extremely Higgsino-like, with very large values of M1 and small mu term.  This means the neutralino NSLP branching fraction into photons in extremely suppressed. Thus the typical NLSP decays are $\chi_0 \rightarrow Z/h$.  In the decoupling limit of the MSSM, the ratio of Higgsino-like NLSP the decay width into Z and Higgs is given by
\be
\Gamma_Z/\Gamma_H= \frac{(N_{13}c\beta-N_{14}s\beta)^2}{(N_{13}c\beta + N_{14}s\beta)^2}\frac{(1-m_Z^2/m_{\chi_0}^2)^4}{(1-m_h^2/m_{\chi_0}^2)^4}
\ee
Where $N_1{j}$ are the Higgsino mixing angles of the neutralino, $m_h/m_Z$ are the Higgs/Z masses and $\beta$ is the Higgs mixing angle.
 Whether the Z of Higgs decay of the NLSP is dominant will be determined by the value of $tan\beta$ and the sign of the mu-term.  Lower values of $tan\beta$ and negative $\mu$-term favor a Higgs decay.  Positive mu-term and large $tan\beta$ favor a Z decay as discussed in \cite{Matchev:1999ft}.  For models such as those discussed here, typical SUSY F terms would  be in the range consistent with possible long lifetime and displaced vertex decay of the NLSP.  Possible sensitivity searches for a the process $\chi_0 \rightarrow Z$ with displaced Z were discussed for the 14 TeV run on LHC in \cite{Meade:2010ji}.

In this scenario, the squarks  undergo two body decay to a quark and chargino/neutralino.  Because the resulting chargino is boosted from the squark decay, its own visible decay products, leptons or jets, may now be quite boosted instead of soft.  In the case of low-energy gauge mediation, the squark decay chain contains in the final state a quark, a lepton or pion, a Higgs or Z, and missing energy trough the process.  Higgs decay is dominated by the two bottom quark process. The squark decays are most likely $\tilde{q} \rightarrow q \chi^{\pm} \rightarrow j + \ell + \nu + \chi_0 \rightarrow j +\ell + \nu + Z/h + G $.  With the most probably final states $3j + \ell  + \SmMET/ j + \ell + 2b + \SmMET $.   Squark pair production contains multiple jets, missing energy, a hard charged particle, and a possible secondary displaced vertex.

The sleptons undergo a decay process very similar to the squarks. Interestingly, the sleptons decay will contain multiple jets.  The process may have many particles in the final state $\tilde{\ell} \rightarrow \ell + \chi_{0} \rightarrow \ell + Z/h + G$.  With the most probably final states $ 2j + \ell  + \SmMET$ and $ 2b+  \ell  + \SmMET $.

\section{Sneutrino LSP}

A final possibility for an LSP or NLSP candidate in these scenarios is the sneutrino. The final state sneutrino, whether or not it is stable or unstable, will appear as missing energy to collider searches.  The sneutrino may either be the LSP or NLSP without effecting the final state topologies of the events.  Below in table 4 I give an example spectrum in which the sneutrino is the LSP or NLSP. I have chosen input values $M_1$ 6250 GeV, $M_2$ 6000 GeV, $M_2$ 8000 GeV and a mu term of 500 GeV with $tan\beta$ equals 3.

\begin{table}[h]
\begin{center}
\begin{tabular}{|c c|c c|c c|c c|}
\hline
neutral gaugino &  mass & chargino & mass & squark & mass & slepton & mass  \\
\hline
 $m_g$  & $6250$ &   &  & $m_{u_l}$ & 807 &  $m_{v_l}$ & 250  \\
 $m_{\chi_4^{0}}$ & $8002$ & $m_{\chi_2^{+}}$ & $6001$  & $m_{d_l}$ & 807 & $m_{e_r}$ & 394 \\
 $m_{\chi_3^{0}}$ & $6001$  &   &  &  &   &  &  \\
 $m_{\chi_2^{0}}$ & $500.4$  & $m_{\chi_1^{+}}$ & $499.5$ &   &  &  & \\
 $m_{\chi_1^{0}}$ & $499$  &   &  & $m_{u_r}$ & $792$ & $m_{e_l}$ & 255  \\
 & & & &  $m_{d_r}$ & $800$  &   &  \\

\hline
\end{tabular}
\label{fields}
\end{center}
\end{table}
Here the three flavors of sneutrino are mass degenerate. I have chosen an intermediate value for the $\mu$ term, and I have chosen to make the sleptons lighter than the charginos and neutralinos.  Here the sneutrinos are the LSP or NLSP. This can be achieved easily in models of gauge mediation if the SU(2) mass parameter contributing to scalars is sufficiently smaller than the U(1) mass parameter. In table 5 I list the most relevant decays and branching fractions for the particles with the sneutrino NLSP spectrum.
\begin{table}[h]
\begin{center}
\begin{tabular}{|c c c| c c c |}
\hline
particle &  decay & b.f.  & particle & decay & b.f. \\
\hline
 $\tilde{u_l}$  & $\chi^{\pm}_1 $d & .78 & $\tilde{t}_1$ & $\chi^{\pm}_1 $ b   & .38\\
 -  & $\chi^{0}_1 $ u & .20 & - & $\chi^{0}_1 $ t  & .51 \\
$\tilde{u_r}$  & $\chi^{0}_1 $u & .96 & $\tilde{t}_2$ & $\chi^{\pm}_1 $ b   & .38\\
 &  &  & - & $\chi^{0}_1 $ t   & .57\\

\hline
 $\tilde{d_l}$  & $\chi^{\pm}_1 $u & .48 & $\tilde{b}_1$ & $\chi^{\pm}_1 $ t   & .13\\
 -  & $\chi^{0}_1 $ d & .49 & - & $\chi^{0}_1$ b  & .41 \\
$\tilde{d_r}$  & $\chi^{0}_1 $d & .96 & - & $\chi^{0}_2$ b  & .45 \\
 &  &  & $\tilde{b}_2$ & $\chi^{\pm}_1 $ t   & .99 \\

\hline

 $\chi^{\pm}_1$ & $\tilde{\nu_{\tau}}$ $\tau$ & .73 & $\tilde{\ell}_R$ & $\tilde{\ell_L}$ $\ell^{\mp}\ell^{\pm}$   & .32 \\
 -& $\nu_{\tau}$ $(\tilde{\tau})_2$ & .16 & - & $\tilde{\nu_{\ell}}$ $\ell\nu_{\ell}$   & .619 \\
 - & $\nu_{\tau}$ $(\tilde{\tau})_1$ & .01 & $\tilde{\ell}_L$ & $\tilde{\nu_{\ell}}$ $jj^{'}$   & .69  \\

 $\chi^{0}_1$ & $\tau^{\mp}$ $\tilde{\tau}^{\pm}$ & .93 & - & $\tilde{\nu_{\ell}}$ $\ell\nu_{\ell}$   & .30 \\
 $\chi^{0}_2$ & $\tau^{\mp}$ $\tilde{\tau}^{\pm}$ & .97 &  &   &  \\

\hline
\end{tabular}
\label{fields}
\end{center}
\end{table}

In this scenario the charged left handed sleptons are fairly close in mass to the sneutrino.  Their decay trajectory is quite simple, a charged slepton decays to its same flavor sneutrino and lepton through a three body process involving an offshell W. The process is $\tilde{\ell}\rightarrow \tilde{\nu} W^{*}$.  The final state signatures are thus $2j + \SmMET$  or $\ell + \SmMET$.  If the mass splitting between left handed slepton and the sneutrino is very small, the decay strongly prefers  the light flavors of leptons.  This has the interesting consequence that a stau will very likely decay without producing a substantial number of taus. Decay of right handed sleptons proceeds through offshell neutralinos and charginos to left handed sleptons plus additional leptons.  For example decay through the off-shell chargino proceeds $\tilde{\ell_R} \rightarrow \nu \chi^{\pm *} \rightarrow \nu \ell^{'} \tilde{\nu}^{'} $.  Interestingly the final state lepton-sneutrino pair may have any flavor.

The slepton decays chains are quite unusual and slepton pair production may have quite an interesting topology.    I show, in figure 7 an illustration of a process whereby left handed sleptons are pair produced.  The final state signature for such a process is most likely soft jets plus missing energy, as the W is quite off shell. This is  a challenging channel for observation of states with electroweak production cross section.  Pair production of left handed charged slepton and sneutrino is similarly hard to observe.  Pair production of right handed sleptons is more interesting in this scenario.  The most likely signature for pair production of charged right handed sleptons is 2 leptons plus missing energy. However, the decay may have 2 leptons of differing flavor.  The three body decay of right handed sleptons with the compressed mass spectrum of sparticles, however, may ensure that the resulting distribution of leptons in pair production events are not very hard and may decrease observability of the system.

The charginos and neutralinos have a fairly simple decay.  Since these particles are Higgsino like, they prefer to decay in chains involving taus.  The light chargino is dominated by the  two body decay process, $\chi^{\pm}\rightarrow \tau^{\pm} \tilde{\nu}$.  The neutralino decays are dominated by decay to stau pairs.  This means that neutralinos  cascade decay through an onshell stau, $\chi_{0}\rightarrow \tau^{\mp} \tilde{\tau}^{\pm}\rightarrow \tau^{\mp} W^{*} \tilde{\nu} \rightarrow \tau^{\mp} j j \tilde{\nu}$.  Since there is an offshell W in the neutralino decay it most often produces jets in the final state, though about a quarter of the time it produces a lepton neutrino pair, once again preferring the light flavors of leptons to the tau.  I also show in figure 7, the decay chains for the light neutralinos and charginos.

\begin{figure}[h]
\centerline{\includegraphics[width=10 cm]{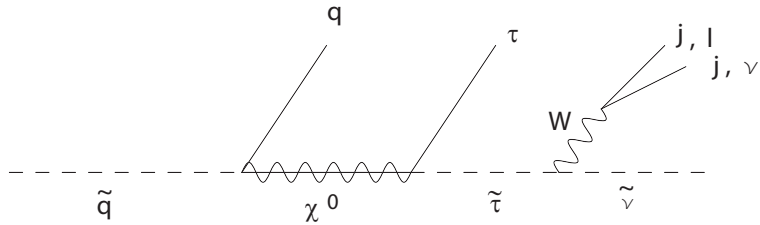}}
\caption{{decay chain.}}
\label{fig:2pgm}
\end{figure}

\begin{figure}[h]
\begin{center}$
\begin{array}{cc}
\includegraphics[width=3.0in]{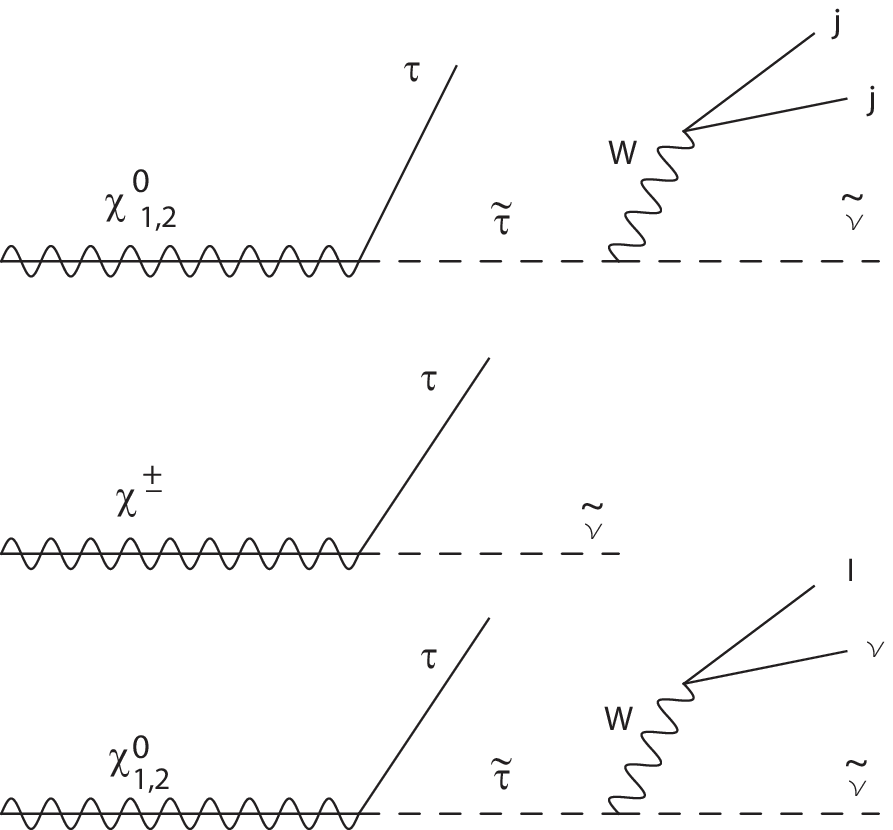} &
\includegraphics[width=3.0in]{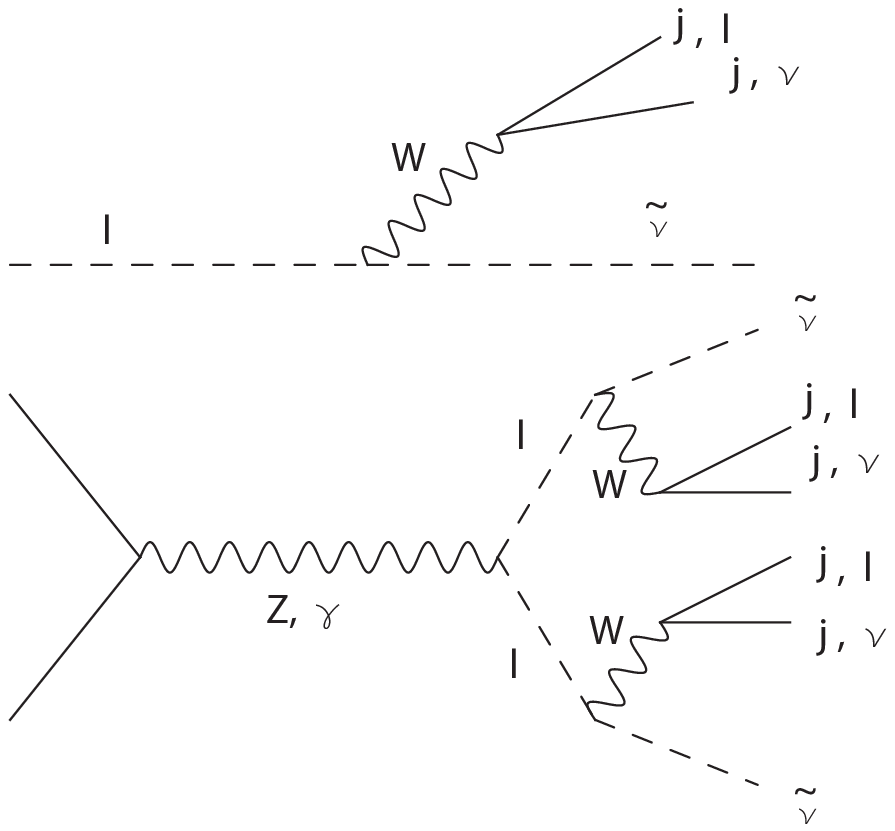}
\end{array}$
\end{center}
\caption{Typical heavy and light flavor squark decay chains in an Antisplit scenario. }
\end{figure}

The squark decay chain in this scenario is quite interesting as it involves a multi-step cascade process.  Again the squarks prefer a two body decay to a quark and onshell gaugino, the gauginos then decay as discussed above to an onshell stau which then decays to the sneutrino LSP through an offshell W.  The typical up-type squark decay chain is then, $\tilde{q_r}\rightarrow  q \chi_{0} \rightarrow q \tau \tilde{\tau}\rightarrow q \tau  \tilde{\nu} W^{*} $.  The offshell W most likely results in jets but result in final state  lepton and neutrino.    Typical quark pair production then produced quite a bizarre signal $\tilde{q} \tilde{q}\rightarrow 6j+2\tau + \SmMET$, or $\tilde{q} \tilde{q}\rightarrow 4j+2\tau+\ell + \SmMET$.  Since it is likely that the decay products of the squarks are boosted, both due to the mass difference between the squarks and the stau, and due to ISR processes, the decay products of the offshell W are likely not too soft to be visible.  It is unclear as to whether current jets plus missing energy searches for squarks are sensitive to this process, and proposing a sensitivity study for this and other processes in the light sneutrino scenario would be quite interesting.


\section{Conclusion}

\begin{table}[h]
\begin{center}
\begin{tabular}{|c|c|c|c|}
\hline
process &  $\tilde{\nu} NLSP$ & Higgsino $NLSP$ & $\tilde{\tau} NLSP$ \\
\hline
 $\tilde{q} \tilde{q}$& & & \\
 & $6j+ 2 \tau + \SmMET$ & $6j+ 4b + \SmMET$ & $2j+ 4 \tau + \SmMET$\\
  & $4j+2 \tau + \ell + \SmMET$ & $4j+ \ell+ 4b + \SmMET$ & $2j+ 2 \tau + \SmMET$\\
 \hline
 $\chi_1^{+}\chi_1^{-}$ & & & \\
  & $2\tau + \SmMET$ & $4j+ 4b + \SmMET$ & $ 2 \tau + \SmMET$\\
  \hline
  $\chi^{0}\chi^{0}$ & & & \\
   & $2\tau +4j+ \SmMET$ & $4b + \SmMET$ & $4 \tau + \SmMET$\\
   & $2 \tau + 2j + \ell + \SmMET$ & & \\
\hline
 $\tilde{\ell} \tilde{\ell}$ & & & \\
 & $4j+\SmMET$ & $2\ell + 4b+\SmMET$  & $2\ell + 2 \tau + \SmMET$ \\
  & $\ell^{'} \ell^{''} + \SmMET$ &   &  $6\ell + 4\tau + \SmMET$ \\

\hline

\hline
\end{tabular}
\label{fields}
\end{center}
\end{table}
In have laid out the phenomenology of a general class as SUSY models where the gauginos are much heavier than the scalars. These models have highly non-standard phenomenology, with particles exhibiting complex decay chains.  Typical of these spectra in general are the fact that the only chargino and neutralinos capable of being produced at collider energies are the Higgsinos, which means the final state topologies of many particle decays large likely to contain highly mass degenerate states and may prefer decay chains with taus.  In addition multiple jets and leptons of varying flavor may occur in sparticle decays.

These models are relatively easy to realize by turning to generalize versions of Gauge Mediation, here I have presented spectra by modifying models in which the scalars are naturally the square root of a loop factor below the gaugino masses.  No doubt there are many other ways to implement this kind of phenomenology. One interesting question is whether other complete mechanisms for SUSY breaking communication can deliver similar Anti-split spectra. For example, one may conceive of building anomaly mediated models in which the scale of $m_{3/2}$ is high, leading to heavy gauginos, while the scalar particles receive extra contributions, say by deflecting the anomaly mediated trajectory with gauge messengers- some kind of General Deflected Anomaly Mediation.  Previous combination of Anomaly Mediation with Dirac Gauginos have lead to interesting spectra and complete, viable Higgs sectors \cite{Carpenter:2005tz}.

Below I summarize phenomenological results with a table of most likely signal topologies for  the most likely decay scenarios for pair production of sparticles in each of the three NLSP scenarios I presented.  Here, since I have picked spectra with modest $tan\beta$ values, I have presented the Higgsino decays as Higgs dominated.

There are many explorations to be made concerning the phenomenology of Anti-split models.  The smoking gun signatures of such a scenario are quite different than what one expects to see in usual scenarios.  Some standard signals are almost completely absent, for example the tri-lepton signature of chargino/neutralino pair production.  While other signatures, such the presence of multiple particle of heavy flavor, are present in many decay chains where they were not before.
Some decay chains in Antisplit scenarios are quite surprising, for example, in the case of sneutrino LSP, the signal for lepton pair production only yields
leptons plus missing energy a few percent of the time, the rest of the time the signal is jets.  Discovery signals in this scenario may be challenging and existing lower mass bounds on this model may be quite weak.

\section{Acknowledgements}
This work was made possible with funds from DOE grant de-sc0013529.

\end{document}